\documentstyle[aps,multicol,epsf]{revtex}
\newcommand{\be}{\begin{equation}}
\newcommand{\ee}{\end{equation}}
\begin{document}
\draft
\title{Nonequilibrium roughening transition by two-species particles\\} 
\author{S.~Park and B.~Kahng \\}
\address{
Department of Physics and Center for Advanced Materials 
and Devices, Konkuk University, Seoul 143-701, Korea \\}
\maketitle
\thispagestyle{empty}
\begin{abstract}
We introduce an interface growth model exhibiting a nonequilibrium 
roughening transition (NRT). 
In the model, particles consist of two species, and deposit or 
evaporate on one dimensional substrate according to a given 
dynamic rule. When the dynamics is limited to occur on 
monolayer, this model has two absorbing states, belonging 
to the directed Ising (DI) universality class. At criticality, 
the density of vacant sites at the bottom layer in the growth model 
decays faster than the DI behavior, however, the dynamic exponent 
is close to the DI value, suggesting that the dynamics is 
related to the DI universality class. We also consider an 
asymmetric version of the growth dynamics, which is according 
to the directed percolation behavior.
\end{abstract}
\begin{multicols}{2}
\narrowtext
Phase transitions in nonequilibrium systems are distinct 
in their physical properties from ones in equilibrium 
systems, in general \cite{review}.  
For example, roughening transition (RT) from a smooth phase 
to a rough phase in one dimension does not occur in thermal 
equilibrium systems \cite{weeks}. 
In nonequilibrium systems, however, there are   
a few examples of exhibiting RT even in one dimension, 
such as the polynuclear growth models \cite{png}, 
the deposition-evaporation model with no evaporation on terrace 
\cite{alon}, and the fungal growth model \cite{fungal}. 
The common feature of the models is that some aspects of the RT 
are related to directed percolation (DP) in 1+1 dimensions \cite{dp}. 
On the other hand, DP behavior occurs in many other 
nonequilibrium systems exhibiting phase transition from active 
to inactive state \cite{dickman}. Typical examples would be 
the monomer-dimer model for the catalytic oxidation of CO 
\cite{zgb}, the contact process \cite{cp}, 
the surface depinning models \cite{depinning}, 
and the branch-annihilation random walks  
with odd numbers of offspring \cite{baw}.  
Recently a lot of efforts have been made to search for other systems 
exhibiting non-DP behavior. 
As a result, a few examples have been found such as the probabilistic 
cellular automata model \cite{grass}, certain kinetic 
Ising model \cite{ki}, the interacting monomer-dimer model \cite{park}, 
modified Domany-Kinzel model \cite{hinrichsen},
and the branch-annihilation random walks with even numbers of offspring (BAWe)
\cite{baw}. 
All these models except for the BAWe have two equivalent absorbing states 
indicating the importance of symmetry in absorbing states to classify 
universality classes.  
Analogous to equilibrium spin models, the non-DP class with 
two equivalent absorbing states is referred to as the directed Ising 
(DI) universality class \cite{hwang}.\\  

Recently the stochastic models [3-5] exhibiting nonequilibrium 
RT (NRT) were introduced, all of which do not possess 
any absorbing state. When absorbing state is not present, 
NRT was understood in terms of the spontaneous symmetry 
breaking of non-conserved order parameter in Refs.[4-5]. 
In this paper, we consider the classification of 
universality classes in NRT having no absorbing state. 
Analogous to phase transition with absorbing state, 
we think that the symmetry lying in dynamic rule 
can replace the role of symmetry in absorbing states, 
and characterizes universality classes. 
In order to confirm this idea,   
we study a simple deposition-evaporation model 
exhibiting NRT, in which particles consist of two species, 
and its dynamic rule is symmetric with respect to particle species. 
We also consider an asymmetric version by breaking the symmetry 
in the dynamic rule. Performing numerical simulations, 
it is found that the symmetric and the asymmetric cases behave 
completely distinctively, confirming that NRT is characterized by 
the symmetry of dynamic rule.
Furthermore, we investigate the classification of 
universality classes for the symmetric and the asymmetric case, 
respectively. For the symmetric case, the density of vacant 
sites at the bottom layer at the threshold of NRT, 
decays faster than the DI behavior, however, 
the dynamic exponent appears close to the DI value, suggesting 
that the growth dynamics is still related to the DI universality 
class. For the asymmetric case, the density behaves well according 
to the DP class. We will discuss physical reason of the different 
behaviors later. \\ 

The stochastic model we introduce is defined as follows. 
We consider an interface dynamics with deposition and 
evaporation processes of particles on one dimensional substrate 
with the periodic boundary condition. 
Particles consist of two species, black and gray colored. 
The dynamics starts from vacuum state, where no particle 
is in the system. First, a site is selected at random. 
Next, either deposition or evaporation of a particle is 
attempted at the selected site with probability $p$ 
($p/2$ for black and $p/2$ for gray particle) 
and $1-p$, respectively, and the attempt is realized under 
the two conditions below. 
First, a restricted solid-on-solid (RSOS) condition is imposed 
such that height difference between nearest neighboring columns 
does not exceed one. 
Second, we consider a ferromagnetic interaction between two nearest 
neighboring particles in the same layer. 
The interaction is attractive (repulsive) when they are 
the same (different) colored. Then a particle with a certain 
color cannot deposit (evaporate) when both of neighboring 
particles in the left and the right sides have different 
(the same) color from its own one as shown in Fig.1. 
Particle can deposit or evaporate when the colors of two 
neighboring particles on each sides are alternative, 
or one (or both) of the neighboring sites is (are) vacant.
Symmetric dynamic rule with respect to particle-color 
generates symmetry in surface configurations; 
every surface configuration has its corresponding configuration 
with replacing one species by the other species as shown in Fig.1.\\ 

In an asymmetric version, one-species particles, say black colored, 
can distinguish the same species from the other species, 
however, gray particles cannot see colors and regard black 
particles as the same species.  
Then a gray particle can deposit (cannot evaporate) even at the site 
where both of two neighboring particles in each sides are black. 
Then the symmetry in the dynamic rule breaks up, and the gray 
flourish more than the black. As an extreme case of the 
asymmetric version, when particles are of single species,      
the model is reduced to the one by Alon $et$ $al$ \cite{alon}, where 
particles cannot evaporate at any site in terrace, 
and only can do at the edge of the terrace. 
The model defined so far is called the growth model hereafter to 
distinguish itself from the model below, called the monolayer 
model, confined in monolayer.\\ 

When the dynamics is restricted on monolayer, so that 
particle cannot deposit on top of particle, 
the model exhibits a phase transition from active 
state to absorbing state.
In this case, vacant site means active site, 
and site occupied by black or gray particle means 
inactive site. There are two types of inactive sites according 
to colors. Absorbing state is formed when the entire system 
is filled with a single species of particles. 
Then there exist two absorbing states which are equivalent 
as long as the dynamic rule is symmetric 
with respect to particle species. Thus, it is expected that 
the phase transition of the monolayer model is in the 
DI universality class. Our monolayer model is similar to 
the generalized contact process proposed by Hinrichsen 
\cite{hinrichsen}, however, our growth model is much simpler 
because it includes one control-parameter rather than two. 
Also, our model is easily generalized into the cases of 
higher symmetry involving $q$-species particles or 
of higher dimensions. 
Moreover, our model might be relevant to describe the 
deposition-evaporation process of Ising particles on surface.\\ 

For the growth model, when $p$ is small, a smooth phase 
is maintained. In this phase, particles form small-sized 
islands which disappear after their short lifetime. 
Thus the growth velocity is zero in the thermodynamic limit. 
As $p$ increases, deposition increases and typical islands grow, 
until, above a critical value $p_c$, islands merge and 
fill new layers completely, giving the interface a finite 
growth velocity. Accordingly, surface exhibits NRT 
from a smooth phase to a rough phase across $p_c$.  
We measure the density of vacant sites $\rho_g(p,t)$ 
at the bottom layer averaged over all runs, where the subscript $g$ 
means the growth model. $\rho_g(p,t)$ is saturated at a 
finite value for $p < p_c$ and decreases to zero exponentially 
for $p > p_c$ in the long time limit as shown in Fig.2(a).  
At criticality, $\rho_g(p_c,t)$ scales 
algebraically as 
\be 
\rho_g(p_c,t) \sim t^{-\beta/\nu_{\parallel}}. 
\ee
We performed Monte Carlo simulations for various system 
sizes $L=10 \sim 1000$, and examined the power-law 
behavior by scanning the probability $p$. 
For the asymmetric case, $p_c$ is estimated as $p_c \approx 0.3796$, 
and $\beta/\nu_{\parallel}\approx 0.16(1)$ is measured, which is  
in good agreement with the DP values, 
$\beta/\nu_{\parallel}(DP)\approx 0.1595$\cite{dickman}.\\ 

For the symmetric case, $p_c$ is estimated as $p_c \approx 0.4480$. 
$\beta/\nu_{\parallel} \approx 0.58(1)$ is measured in the long 
time limit, which is much deviated from the DI value, 
$\beta/\nu_{\parallel}(DI)\approx 0.27 \sim 0.29$ [12-17]. 
We think that this discrepancy may come from the suppression-effect 
by particle on upper layer; 
For example, particle A in Fig.1 can evaporate 
in the monolayer model because one of neighboring 
particles has different color, however, 
it cannot evaporate in the growth model because of another 
particle on top of it.
On the other hand, for the asymmetric case, particle A 
cannot evaporate even in the monolayer model, because two 
particles on each side are regarded as the same species. 
Therefore, the suppression-effect appears strongly (weakly) 
for the symmetric (asymmetric) case. 
When number of particles on upper layers is relatively small 
for $p \ll p_c$, or in the short time regime for $p=p_c$, 
the suppression-effect is relatively weak. 
Thus, $\rho_g(p_c,t)$ decays according to the DI behavior  
in the short time regime as indicated by dashed line in Fig.2(a). 
However, in the long time limit at $p_c$, evaporation is 
suppressed by particles on upper layer, and $\rho_g(p_c,t)$ 
decays as $\sim t^{-0.58}$, faster than the DI behavior. 
Note that the value $\approx 0.58$ is almost 
double to the DI value.\\ 
 
We consider the probability $S(t)$ that the system contains 
at least one vacant site on the bottom layer, and other sites are 
occupied by any species of particles. Note that $S(t)$ is 
different from the survival probability $P(t)$ in the 
monolayer model, in that the probability $S(t)$ ($P(t)$) counts 
the configurations that the bottom layer is filled with particles 
of any species (a single species). As can be seen in Fig.2(b), 
there exists a characteristic time $\tau_g$ such 
that for $t < \tau_g$, $S(t)=1$, and for $t> \tau_g$, 
$S(t)$ decays in the same way as does $\rho_g$. 
Thus the density of vacant sites $\rho_g^{(s)}(p_c,t)$, averaged 
over samples with at least one vacant site, 
decays as Eq.(1) until $\tau_g$, and is finite 
beyond $\tau_g$ as shown in Fig.2(c). 
The steady state value of $\rho_g^{(s)}(p_c)$ depends 
on system size $L$ as 
$\rho_g^{(s)}(q_c) \sim L^{-\beta/\nu_{\perp}}$. 
We obtained $\beta/\nu_{\perp}\approx 0.95(1)$, 
which is almost double to the DI value, $\approx 0.5$. 
However, the value of the dynamic exponent we measured 
$\nu_{\parallel}/\nu_{\perp}\approx 1.64(2)$, 
describing the characteristic time via 
$\tau_g \sim L^{\nu_{\parallel}/\nu_{\perp}}$, 
is close to the DI value, $\approx 1.66 \sim 1.75$.  
We also examine the steady-state density $\rho_g(p,\infty)$ 
for various $\epsilon \equiv (p_c-p) > 0$, which 
may be written as $\rho_g(p,\infty) \sim \epsilon^{\beta}$. 
As shown in Fig.2(d), the data do not fit well to 
a straight line for small $\epsilon$, but are likely 
to approach a line asymptotically with slope $\beta\approx 0.88$, 
the DI value, for large $\epsilon$, being far from $p_c$. 
Accordingly, the dynamics of the growth model seems to 
be related to the DI class, but the correlation length 
exponents $\nu_{\parallel}$ and $\nu_{\perp}$ seems to be 
halves of the DI values.\\ 

In the rough phase, surface grows with finite velocity, 
which behaves as $v \sim (q-q_c)^{y}$. 
The velocity would be characterized by the 
inverse of the characteristic 
time $\tau_g$ to complete one layer, $v \sim a/\tau_g$, 
where $a$ is lattice constant.  
If $\tau_g$ is regarded as the characteristic time 
$\tau_m$ for the monolayer model to reach the absorbing state, 
$\tau_m \sim (q-q_c)^{-{\bar \nu}_{\parallel}}$, 
where the symbol of bar means the monolayer model, 
then the velocity exponent $y$ would be equal 
to ${\bar \nu}_{\parallel}$. 
This assumption is confirmed in the asymmetric case by that  
the value measured, $y \approx 1.67(8)$, is close to 
the DP value, $\nu_{\parallel}(DP)\approx 1.73$. 
However, for the symmetric case, we measured 
$2y \approx 2.50(6)$, which is deviated from  
$\nu_{\parallel}(DI)\approx 3.17\sim 3.25$ \cite{park}. 
This discrepancy may result from that the suppression-effect 
appears much strongly in rough phase.\\  

Next, we also consider the surface fluctuation width, 
$W^2(L,t)={1\over L}\sum_i h_i^2(t)- \big({1\over L}\sum_i h_i(t)
\big)^2$,
where $h_i(t)$ denotes height at site $i$ at time $t$. 
The interface width behaves as 
\be 
W^2(L,t) \sim \cases{
t^{2\zeta/z}, 
& for $t \ll L^{1/z}$, \cr 
(-\epsilon)^{\chi}L^{2\zeta}, 
& for $t \gg L^{1/z},$ \cr} 
\ee
where $\zeta$ and $z$ are called the roughness 
and the dynamic exponent, respectively. 
$\chi$ is the exponent describing the vanishing of the 
roughness as the transition is approached. The 
exponent $\chi$ is measured as $\chi \approx 0.34(1)$ 
($\approx 0.89(4)$) for the symmetric (asymmetric) case 
as shown in Fig.3(b).  
The asymmetric value is close to the value 
$0.92$ obtained from the single-species model \cite{alon}. 
The roughness exponent $\zeta$ is measured to be $\zeta  
\approx 0.50(1)$, which is very close to the value of the 
Edwards-Wilkinson (EW) universality class \cite{ew} and the 
Kardar-Parisi-Zhang (KPZ) universality class \cite{kpz} 
in one dimension. The value of $\zeta \approx 0.5$ is also obtained 
from the height-height correlation function, 
$C^2(r)=<{(h(r)-h(0))^2}>\sim r^{2\zeta}$ in the long time 
limit as shown in Fig.3(c). 
The growth exponent $\zeta/z$ exhibits a broad range of 
numerical values from $1/4$ to $1/3$ as the 
probability $p$ increases as shown in Fig.3(d). 
Since the velocity for $p > p_c$ is nonzero, 
the EW universality class for surface growth has to be excluded. 
Nevertheless, since the velocity is extremely small near $p_c$, 
the EW behavior appears numerically for finite system, 
but in the thermodynamic limit, surface growth for 
$p > p_c$ is expected to belong to the KPZ class. 
At $p_c$, the roughness of surface 
exhibits the marginal behavior, $W^2 \sim \log t$ 
for $t \ll L^{1/z}$ and $W^2 \sim \log L$ for $t \gg L^{1/z}$.\\

Next, we perform numerical simulations for the monolayer 
model. We start with two distinct initial configurations.
First, the initial configuration includes one active site, 
so that all sites are occupied by a single species particles 
except one empty site. Then number of active sites 
increases with increasing time.  
We measure the survival probability $P(t)$ (the probability 
that the system is still active at time $t$), the 
density of active sites $\rho_m(t)$ averaged over all runs, 
and the mean-square-distance of spreading of the active 
region $R^2(t)$ averaged over surviving runs. 
At criticality, ${\bar p}_c \approx 0.7485$, 
the values of these quantities scale algebraically 
in the long time limit as 
$P(t)\sim t^{-\bar \delta}$, $\rho_m(t) \sim t^{\bar \eta}$, 
and $R^2(t) \sim t^{\bar z}$. We measured the exponents as 
$\bar \delta \approx 0.287(1)$, $\bar \eta \approx 0.000$, and 
$\bar z \approx 1.141(3)$, all of which are in good agreement 
with those in the DI universality class \cite{exact}.
Next, we start with another initial configuration 
where all sites are empty. This initial configuration is considered 
for the purpose of comparing with the growth model. 
In this case, number of active sites decreases 
with increasing time. 
The density of active sites behaves as 
$\rho'_m(t) \sim t^{-\bar{\eta'}}$ 
and the survival probability does as $P'(t) \sim t^{-{\bar \delta'}}$ 
with the exponents $\bar{\eta'} \approx 0.279(1)$ 
and $\bar \delta' \approx 0.000$, respectively. 
Even though the two different initial configurations 
yield the different values of the exponents,
their sum ${\bar \delta}+{\bar \eta}$ seems to be 
the same as ${\bar \delta'}+{\bar \eta'}$, 
describing the growth of the number of kinks over 
surviving samples in the interacting monomer-dimer model \cite{park}.  
Detailed numerical results for the growth model and the monolayer 
model will be published elsewhere \cite{next}.\\

In summary, we have introduced an interface growth model 
with deposition and evaporation of two species particles,  
exhibiting a nonequilibrium roughening transition (NRT) from 
a smooth phase to a rough phase as deposition-attempt 
probability $p$ varies. The dynamic rule of the deposition 
and evaporation process is assigned symmetrically and 
asymmetrically with respect to particle species, respectively. 
For the two cases, NRTs behave distinctively from each other, 
suggesting that the universality class of NRT is 
characterized by the symmetry lying in dynamic rule. 
For the symmetric case, some features of NRT are related to 
the DI class, but the correlation length exponents $\nu_{\parallel}$ 
and $\nu_{\perp}$ in the growth models are smaller by halves 
than those of the monolayer model in the DI class, 
which may result from the strong suppression-effect by particles 
on upper layer. For the asymmetric case, 
however, NRT are described by the DP behavior.\ 

BK wish to thank H. Park and H. Hinrichsen for helpful discussions. 
This work was supported by the Ministry of Education, 
Korea (97-2409).  


\begin{figure}
\centerline{\epsfxsize=7.3cm \epsfbox{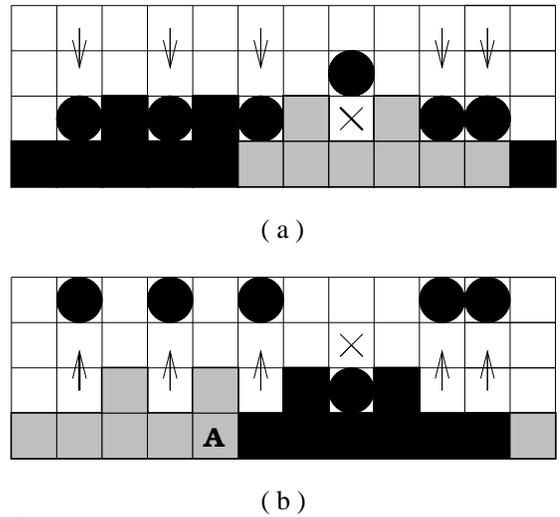}}
\caption{The dynamic rule for deposition (a) and 
for evaporation (b) for the symmetric case of the growth model.} 
\label{fig1}
\end{figure}

\begin{figure}
\centerline{\epsfxsize=8.3cm \epsfbox{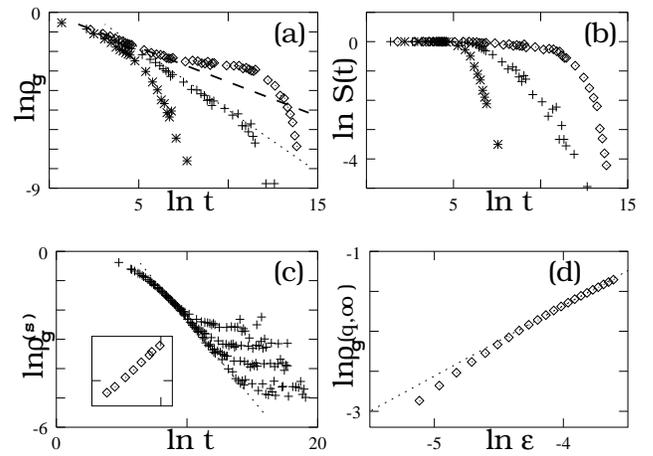}}
\caption{(a) Double logarithmic plot (DLP) of $\rho_g(p,t)$ 
versus time $t$ for probabilities $p=0.4410$ (top), 
$0.4480 (=p_c)$ and $0.4600$ (bottom).
The data are obtained for $L=100$, averaged over more than 
500 configurations. The dashed and the dotted lines have 
slope 0.276 and 0.589, respectively, drawn for the 
eye.   
(b) DLP of $S(t)$ versus time $t$ for the same cases as (a). 
(c) DLP of $\rho_g^{(s)}(p_c,t)$ versus $t$ 
for different system sizes $L=50$(top),100, 200 and $500$. 
The dotted line has slope 0.589, drawn for the eye.  
Inset: DLP of $\tau_g$ versus $L$ at $p_c$. 
(d) DLP of $\rho_g(p,\infty)$ versus $\epsilon$. 
The dashed line has slope 0.88, drawn for the eye. 
All plots (a-d) are for the symmetric case of the growth model.}
\label{fig2}
\end{figure}

\begin{figure}
\centerline{\epsfxsize=8.3cm \epsfbox{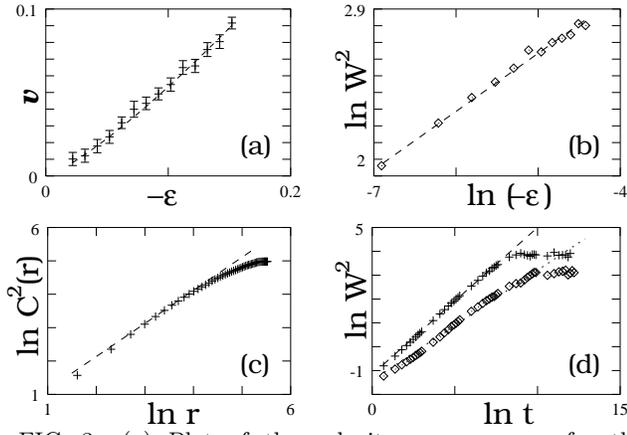}}
\caption{(a) Plot of the velocity $v$ versus $-\epsilon$ for 
the symmetric case of the growth model. The dashed line 
$v=0.95*(-\epsilon)^{1.25}$ was obtained by a least-square-fit. 
(b) DLP of $W^2$ versus $-\epsilon$ to measure the exponent $\chi$. 
The dotted line, a guideline to the eye, has slope $0.34$.
(c) DLP of the height-height correlation function versus distance $r$ 
at $p=0.9$. 
The data are well fit to a straight line with slope $2\zeta=1$.
(d) DLP of $W^2$ versus $t$ at probabilities $p=0.5$ and $0.9$(top). 
The dotted and dashed lines have slopes $0.49$ and $0.62$(top) to guide 
to the eye. 
All plots (a-d) are for the symmetric case of the growth model 
with $L=500$, and the data are averaged over 500 configurations.} 
\label{fig3}
\end{figure}

\end{multicols}

\begin{thebibliography}{99}
\bibitem{review} For a review, see, V. Privman, 
{\it Nonequilibrium phase transitions in lattice models} 
(Cambridge University, Cambridge, 1996).  
\bibitem{weeks} J.~D.~Weeks, in {\it Ordering in Strongly
Fluctuating Condensed Matter Systems}, edited by T. Riste (Plenum,
New York, 1980).
\bibitem{png} J. Kert\'esz and D.E. Wolf, {\rm Phys. Rev. Lett.} 
{\bf 62}, 2571 (1989).
\bibitem{alon} U. Alon, M.R. Evans, H. Hinrichsen and D. Mukamel, 
{\rm Phys. Rev. Lett.} {\bf 76,} 2746 (1996); {\rm Phys. Rev. E} 
{\bf 57}, 4997 (1998).
\bibitem{fungal} J. M. L\'opez and H. J. Jensen, (cond-mat/9803171).
\bibitem{dp} G. Deutscher, R. Zallen and J. Adler, 
{\it Percolation Structures and Processes} Ann. Isr. Phys. Soc.5 
(Adam Hilger, Bristol, 1983).
\bibitem{dickman} J. Marro and R. Dickman, in Ref.[1]. 
\bibitem{zgb} R.M. Ziff, E. Gulari and Y. Barshad, 
{\rm Phys. Rev. Lett.} {\bf 56,} 2553 (1985). 
\bibitem{cp} T.E. Harris, {\rm Ann. Prob.} {\bf 2}, 969 (1974); 
T.M. Liggett, {\it Interacting Particle Systems} 
(Spinger-Verlag, New York,1985)   
\bibitem{depinning}
A.-L. Barab\'asi and H. E. Stanley, {\it Fractal Concepts in Surface Growth}
(Cambridge University Press, Cambridge, England, 1995);
D. Kim, H. Park and B. Kahng, {\it Dynamics of Fluctuating Interfaces and
Related Phenomena} (World Scientific, Singapore, 1997).
\bibitem{baw} H. Takayasu and A. Yu Tretyakov, {\rm Phys. Rev. Lett.} 
{\bf 68}, 3060 (1992).
\bibitem{grass} P. Grassberger, F. Krause and T. von der Twer, 
{\rm J. Phys. A} {\bf 17}, L105 (1984). 
\bibitem{ki} N. Menyh\'ard, {\rm J. Phys. A} {\bf 27,} 6139 (1994).  
\bibitem{park} M.H. Kim and H. Park, {\rm Phys. Rev. Lett.} {\bf 73,} 2579 
(1994); H. Park and H. Park, {\rm Physica A} {\bf 221,} 97 (1995).
\bibitem{hinrichsen} H. Hinrichsen, {\rm Phys. Rev. E} {\bf 55,} 219 (1997).
\bibitem{hwang} W. Hwang, S. Kwon, H. Park and H. Park, 
{\rm Phys. Rev. E} {\bf 57}, 6438 (1998).
\bibitem{exact} N.Inui and A. Yu. Tretyakov, {\rm Phys. Rev. Lett.} {\bf 80,} 
5148 (1998).
\bibitem{ew} S.F. Edwards and D.R. Wilkinson, {\rm Proc. R. Soc. Lond. A} 
{\bf 381,} 17 (1982). 
\bibitem{kpz} M. Kardar, G. Parisi and Y.C. Zhang, {\rm Phys. Rev. Lett.} 
{\bf 56,} 889 (1986).  
\bibitem{next} S. Park and B. Kahng, (unpublished).
\end{thebibliography}
\end{document}